\documentclass[amstex,aps,showpacs]{revtex4}%
\usepackage{graphicx}
\usepackage{amsmath}
\usepackage{bm}%
\usepackage{amsfonts}%
\usepackage{amssymb}

\begin{document}
\title
{Quantum and semiclassical phase functions for the quantization of
symmetric oscillators}
\author{A.\ Matzkin and M. Lombardi }
\affiliation{Laboratoire de Spectrom\'{e}trie physique (CNRS
Unit\'{e} 5588), Universit\'{e} Joseph-Fourier Grenoble-I, BP 87,
38402 Saint-Martin, France}

\begin{abstract}
We investigate symmetric oscillators, and in particular their
quantization, by employing semiclassical and quantum phase
functions introduced in the context of Liouville-Green
transformations of the Schr\"{o}dinger equation. For anharmonic
oscillators, first order semiclassical quantization is seldom
accurate and the higher order expansions eventually break down
given the asymptotic nature of the series. A quantum phase that
allows in principle to retrieve the exact quantum mechanical
quantization condition and wavefunctions is given along with an
iterative scheme to compute it. The arbitrariness surrounding
quantum phase functions is lifted by supplementing the phase with
boundary conditions involving high order semiclassical expansions.
This allows to extend the definition of oscillation numbers, that
determine the quantization of the harmonic oscillator, to the
anharmonic case. Several illustrations involving homogeneous as
well as coupling constant dependant anharmonic oscillators are
given.
\\

Journal Reference: J. Phys. A 38, 6211 (2005).
\end{abstract}

\pacs{03.65.Sq,03.65.Ca,03.65.Ge,02.60.Lj,03.65.Vf}

 \maketitle

\section{Introduction}

The harmonic oscillator is quite often employed as a prototype for
illustrating numerous phenomena in different areas of quantum mechanics. It is
somewhat infortunate, since the harmonic oscillator is all but atypical, even
within the family of symmetric oscillators. In particular the quantization of
symmetric oscillators is radically different in the harmonic and anharmonic
cases: as is very well-known, the first order semiclassical quantization
condition, given by
\begin{equation}
S(t_{2},E)-S(t_{1},E)=\pi\hbar(n+\frac{1}{2}) \label{e1}%
\end{equation}
where $S$ is the classical action, $t_{1}$ and $t_{2}$ are the
turning points and $n$ is the level integer for the energy $E$, is
exact for the harmonic oscillator, but fails to capture even the
most elementary aspects in the anharmonic case. For example, for
homogeneous potentials $x^{2m}$ ($m\neq1$), Eq. (\ref{e1})
predicts that the ground state energy should decrease when $m$
increases, whereas the exact quantum ground state levels behave
the other way round ($E$ increases with $m$). It is therefore no
surprise that considerable theoretical and computational efforts
have been made to investigate these systems.\ On the one hand,
symmetric oscillators are among the simplest nonsolvable systems
\cite{note1}. On the other hand they are employed in many branches
of quantum physics, ranging from molecular vibrations to simple
models of quantum field theories.

The quantization of symmetric oscillators has largely focused on
the accuracy of the standard semiclassical expansion. Eq.
(\ref{e1}) can usually be improved by going to higher order in
$\hbar$, as will be briefly recalled in Sec. 2. But the asymptotic
series is factorially divergent, so at best the expansion must be
truncated, although the accuracy rapidly increases with $n$
\cite{bender etal77}. The possibility of employing resummation
techniques and resurgence analysis was also extensively
investigated, especially on the quartic oscillator
\cite{voros83}.\ Proofs regarding the convergence of resurgence
schemes do exist in some particular cases \cite{pham??}, though it
seems hardly realistic to base practical calculations on such
schemes. Alternatively, it was pointed out \cite{balian voros77}
that symmetric oscillators have turning points away from the real
line and that including the subdominant contributions from these
complex trajectories (where space-time coordinates become complex)
accounts for most of the discrepancies between the best
semiclassical estimate and the exact quantum result. It was even
recently suggested that having particles moving along trajectories
in the complex plane should be regarded as a real physical feature
\cite{chebotarev99}. We also mention the importance of anharmonic
oscillators of the form $x^{2}+\lambda x^{2m}$ in the
investigation of quantum perturbation theories \cite{bender wu69}.

Thus, even for simple systems such as symmetric oscillators with a
single minimum, quantization is far from being understood. In the
present work we analyze the quantization of symmetric oscillators
(determination of bound states in potential wells with a single
minimum) from the perspective of an exact quantum phase. The phase
-- a real function -- is obtained from a Liouville-Green
transformation of the Schrodinger equation. It achieves exact
quantization and allows to retrieve the exact quantum mechanical
wavefunction. It also extends to a general symmetric oscillator
certain quantities that only exist for the harmonic oscillator,
such as the oscillation number. Quantum phase functions suffer
from ambiguities in their definitions; the idea to be implemented
here is to obtain a phase function that will be close, in a sense
to be precised below, to the predivergent semiclassical expansion.
We will also give a simple and efficient numerical procedure to
construct the phase function, achieving quantization with an
a-priori unlimited precision. We will first recall the main
semiclassical quantization schemes (Sec.\ 2). The standard
semiclassical approximation is the most common one, but it is
well-known (though too often overlooked) that $\hbar$ expansions
can be carried out by employing alternative schemes, based on
different transformation functions. We will then introduce the
quantum phase by making a specific transformation that will be
justified, in particular by its relation to the semiclassical
phase (Sec. 3). In Sec. 4 we will illustrate the roles of quantum
and semiclassical phases in the harmonic oscillator case: this is
a solvable problem that will further allow us to gain insight in
the meaning of the quantum phase by working with the analytical
solutions. We will in particular introduce the notion of optimal
quantum and semiclassical phase functions by linking them to the
exact and semiclassical oscillation number. After giving the
numerical procedure to compute the quantum phase (Sec. 5), we will
deal with the quantum phase investigation of anharmonic symmetric
oscillators in Sec. 6.\ We will first insist on the pure quartic
oscillator, which is the one that has been the most heavily
studied. We will also illustrate some properties of sextic and
octic oscillators as well as the behaviour as a function of
$\lambda$ for a harmonic oscillator perturbed by a $\lambda
x^{10}$ anharmonicity. Our closing comments and our conclusions
will be given in Sec.\ 7.

\section{Semiclassical quantization schemes}

\subsection{Standard complex plane quantization}

The standard semiclassical approach in one dimension -- the WKB scheme -- is
based on a Riccati transform of the wavefunction followed by an $\hbar$
expansion. The first step is to write the wavefunction $\psi(x)$ as%
\begin{equation}
\psi(x)=\exp\left[  i\zeta(x)/\hbar\right]  \label{e3}%
\end{equation}
thereby transforming the Schrodinger equation for $\psi$,%
\begin{equation}
\hbar^{2}\partial_{x}^{2}\psi(x)+p^{2}(x)\psi(x)=0, \label{e4}%
\end{equation}
into a Riccati-type equation for the phase $\zeta$. By the argument principle,
the logarithmic derivative of $\psi$ integrated along a contour encircling the
$n$ zeros of $\psi$ yields the quantization condition (see eg \cite{slavyanov}%
)%
\begin{equation}
\oint\partial_{z}\zeta(z)dz=2\pi\hbar n. \label{e5}%
\end{equation}
Substituting the expansion
\begin{equation}
\zeta(z)=\sum_{k=0}^{\infty}\left(  -i\hbar\right)  ^{k}\zeta_{k}(z)
\label{e7}%
\end{equation}
into the Riccati equation and equating like powers of $\hbar$
yields
$\zeta_{0}(x)=S(x)$ and the recurrence relation%
\begin{equation}
\partial_{z}^{2}\zeta_{k-1}+\sum_{j=0}^{k}\partial_{z}\zeta_{k-j}\partial
_{z}\zeta_{j}=0. \label{e9}%
\end{equation}
Eq. (\ref{e5}) was obtained by Dunham \cite{dunham32}. The
solutions of Eq. (\ref{e9}) are then plugged into the \ asymptotic
expansion (\ref{e7}) which is substitued in Eq. (\ref{e5}).\ Note
that the contour in Eq. (\ref{e5}) must now enclose the turning
points \cite{froman77} since the solutions $\zeta_{k}$ are
singular at the turning points. The first odd term
$\partial_{z}\zeta _{1}(z)$ integrated along the contour yields
$i\pi$ whereas all the other odd terms are total derivatives and
therefore do not contribute \cite{bender
etal77}. Eq. (\ref{e5}) thus takes the form%
\begin{equation}
\oint\sum_{k=0}^{k_{\max}=\infty}(i\hbar)^{2k}\partial_{z}\zeta_{2k}%
(z,E)dz=2\pi\hbar(n+\frac{1}{2}). \label{e10}%
\end{equation}

Except in specific cases (such as the quartic oscillator for which
the integrals are known analytically), the integrals in Eq.
(\ref{e10}) are taken on the real line and must be regularized
since the functions $\zeta_{2k}^{\prime}$ contain nonintegrable
singularities at the turning points \cite{tutik00}. Finally
inverting Eq. (\ref{e10}) for a finite value of $k_{\max}$ yields
the quantized eigenvalue $E$. As for any asymptotic series
\cite{nayfeh} the general trend as $k_{\max}$ is increased is to
obtain better approximations for the first few terms but quickly
the series diverge (examples will be given below).

\subsection{Alternative asymptotic quantization schemes}

More general asymptotic expansions are readily obtained by
employing alternatives to the transformation
(\ref{e3}).$\;$Different asymptotic ansatzes can be found in Refs.
\cite{slavyanov,olver}. The most useful form is based
on the Liouville-Green transformation whereby the wavefunction is written as%
\begin{equation}
\psi(x)=u(x)w\left(  \xi(x)\right)  , \label{e12}%
\end{equation}
where $u$ and $w$ are two arbitrary (but sufficiently smooth) functions and
$\xi$ appears as a new dependent variable. This transformation can be
restricted by requiring two linearly independent solutions of the
Schr\"{o}dinger equation $\psi_{1}$ and $\psi_{2}$ to have exactly the same
form (\ref{e12}) with two different functions $w_{1}$ and $w_{2}$. Recalling
that the Wronskian $\mathcal{W}[\psi_{1}$,$\psi_{2}]$ is a constant, we are
led to the transformation%
\begin{equation}
\psi(x)=\left(  \partial_{x}\xi(x)\right)  ^{-1/2}w(\xi(x)). \label{e13}%
\end{equation}
Assume $w(\xi)$ fulfills the equation%
\begin{equation}
\hbar^{2}\partial_{\xi}^{2}w(\xi)+R(\xi)w(\xi)=0, \label{e14}%
\end{equation}
where the choice of the unspecified but smooth function $R(\xi)$ determines
the choice of $w$. $\xi$ then obeys
\begin{equation}
R(\xi)\left(  \partial_{x}\xi\right)  ^{2}-p^{2}(x)+\frac{\hbar^{2}}%
{2}\left\langle \xi;x\right\rangle =0, \label{e15}%
\end{equation}
where $\left\langle \xi;x\right\rangle \equiv\partial_{x}^{3}\xi/\partial
_{x}\xi-\frac{3}{2}(\partial_{x}^{2}\xi/\partial_{x}\xi)^{2}$ denotes the
Schwartzian derivative. The procedure is now to employ an asymptotic expansion
for $\xi,$%
\begin{equation}
\xi(x)=\sum_{k=0}^{\infty}\xi_{2k}(x)\hbar^{2k} \label{e16}%
\end{equation}
which is substituted into Eq. (\ref{e15}) to obtain a recurrence system with
the first term being found from%
\begin{equation}
R(\xi_{0})\left[  \partial_{x}\xi_{0}(x)\right]  ^{2}=p^{2}(x). \label{e18}%
\end{equation}
Note that the odd terms in the $\hbar$ expansion (\ref{e16}) are redundant.

The advantage of the present scheme relative to the standard
semiclassical treatment recalled above is twofold. First the
choice of $R$, which determines both $w(x)$ and $\xi(x)$ allows a
greater flexibility. In particular it allows to construct
wavefunctions that are well defined on the entire real line, so
that although the preceding analysis is valid for both complex and
real variables, it is possible to work with real quantities only.
This is of course not necessary, since the so-called 'phase
integral method' \cite{froman02} basically amounts to choosing the
exponential function for $w$ but keeping complex values for $\xi$,
therefore differing from the standard method by requiring Eq.
(\ref{e13}) to be enforced at each order. But as we shall argue
below, working with real functions brings in a second advantage,
which is of conceptual nature, namely that it allows to better
understand the classical limit, or conversely to better follow
what the classical quantities becomes in the quantum domain.

\section{Quantum and semiclassical phase functions}

\subsection{Choice of the phase: carrier functions}

It is apparent that putting the wavefunction in the form given by Eq.
(\ref{e13}) is tantamount to undertaking an amplitude-phase decomposition,
where $\xi(x)$ appears as a phase and
\begin{equation}
\alpha(x)=\left(  \partial_{x}\xi(x)\right)  ^{-1/2} \label{e20}%
\end{equation}
appears as an amplitude function. Indeed Eq. (\ref{e20}) simply
represents the one-dimensional continuity equation. $w$ represents
here a 'carrier' function, as it carries the phase. It is clear
that defining $\xi(x)$ as a quantum phase is ambiguous: first, the
carrier function needs to be specified (thereby deducing $R(\xi)$)
and second the boundary conditions of the third order nonlinear
equation (\ref{e15}) need to be given.\ From a purely internal
quantum mechanical viewpoint, it would appear at first sight that
the choice of $\xi(x)$ is meaningless, insofar as the quantum
mechanical quantities are insensitive to a specific manner of
cutting the wavefunction. From a semiclassical point of view
things are different, since different choices of $\xi(x)$ yield a
different first order semiclassical phase $\xi_{0}(x).$ From Eq.
(\ref{e18}) it is straightforward to see that the only choice
leading to $\xi_{0}(x)=\left|  S(x)\right|  $ corresponds to
$\left|  R(\xi_{0})\right| =1$, corresponding to exponential or
circular carrier functions. Note that in that case, the terms
$\xi_{2k}(x)$ are identical in absolute value to the even terms of
the expansion (\ref{e7}), thereby leading to the same quantization
condition; the odd terms are different, although in both cases
they are singular at the turning points. Of course this is not the
case if Eq. (\ref{e15}) is to be solved exactly, since the exact
solutions are well defined on the entire real line

As an example of a semiclassical phase different from the classical action,
consider taking $R(\xi)=\xi$, corresponding to Airy carrier functions. This
case has a practical interest because it yields an asymptotic expansion not
diverging at the turning points. Given the symmetry of the oscillators, it is
convenient to restrict the analysis to the half line, say $]-\infty,0].$ The
first order solution can be written in the handy form
\begin{equation}
\xi_{0}(x)=\mp\left(  \pm\frac{3}{2}\right)  ^{2/3}\left[  \int_{t_{1}}%
^{x}\left|  p(x^{\prime})\right|  dx^{\prime}\right]  ^{2/3}, \label{e21}%
\end{equation}
where the signs change when $x$ crosses $t_{1}$ due to the Stokes
phenomenon. Eq. (\ref{e21}) appears in the context of comparison
equations when $p(x)$ admits one single turning point; see eg Ref.
\cite{slavyanov}, in particular Sec. II.3, where it is shown that
$\xi_{0}(x)$ as well as the higher-order solutions $\xi_{2k}(x)$
are well-behaved in the vicinity of $t_{1}$.\ From a purely
semiclassical perspective it is often advantageous to employ in
the same problem different comparison equations to expand the
solution in the neighbourhood where the relevant approximation
holds best \cite{alvarez00}, leading to the matching of different
phase functions.  Whether $\xi(x)$ can be called a 'phase' or not
is a question of terminology.\ But it is clear that the choice of
$R(\xi)$ gives different carrier functions and different
asymptotic $\hbar$ expansions for $\xi$. What is interesting in
the present context is that the phase corresponding to the
standard semiclassical choice $R(\xi)=1$ can be obtained in terms
of other phase functions, opening the possibility of performing
any asymptotic expansion for the standard semiclassical phase.
This is done by recasting the phase equation in terms of Ermakov
systems.

\subsection{Choice of the phase: boundary conditions and Ermakov systems}

Uncoupled Ermakov systems relate solutions of the linear differential Eq.
(\ref{e4}) and of a nonlinear equation for $\alpha(x)$ similar to Eq.
(\ref{e15}) when $\left|  R(\xi)\right|  =1$,%
\begin{equation}
\hbar^{2}\partial_{x}^{2}\alpha(x)+p^{2}(x)\alpha(x)=\alpha^{-3}(x).
\label{e23}%
\end{equation}
Eq. (\ref{e23}) was employed very early in a quantum mechanical
context \cite{milne-young-wheeler}, but the connection with
Ermakov systems, based on the nonlinear superposition principle
\cite{ray-reid}, is quite recent (see
\cite{matzkin00,matzkin01,thylwe05} and Refs. therein). The only
results we shall need are the following. Assume $R(\xi)=1$ (ie,
$\sin$ or $\cos$ carrier functions) and denote the phase in this
case by $\sigma(x)\equiv\xi(x)$. Let $\psi_{1}(x)$ be a solution
of Eq. (\ref{e4}) regular at $-\infty$ and $\psi_{2}(x)$ be a
solution regular at $+\infty$ (given the symmetry, we can take
here $\psi_{2}(x)=\psi_{1}(-x)$) and
$W=\mathcal{W}[\psi_{1},\psi_{2}]$
their Wronskian. Eq. (\ref{e13}) can be written as%
\begin{equation}
\psi_{1}(x)=\sqrt{2I}\alpha(x)\sin\sigma(x) \label{e24}%
\end{equation}
where $I$ is the Ermakov invariant. Using $\sigma(-x)=\sigma(\infty
)-\sigma(x)$ and $\sigma(0)=\sigma(\infty)/2$ we also have the identity%
\begin{equation}
W^{2}=4I^{2}\sin^{2}\left[  2\sigma(0)\right]  . \label{e25}%
\end{equation}
A solution of Eq. (\ref{e4}) independent from $\psi_{1}(x)$ and lagging
$\pi/2$ out of phase is given by%
\begin{equation}
\psi_{3}(x;I,c)=2I\left(  \frac{\psi_{2}(x)}{W}-c\psi_{1}(x)\right)
\label{e26}%
\end{equation}
where $c$ is an arbitrary constant. $\psi_{3}$ is in general irregular at
$\pm\infty$.\ From these considerations, it follows that%
\begin{equation}
\tan\sigma(x)=\frac{\psi_{1}(x)}{\psi_{3}(x;I,c)}. \label{e27}%
\end{equation}
We have emphasized the dependence of $\sigma (x)$ on $I$ and $c$,
the form taken in the present context by the two boundary
conditions left from Eq. (\ref{e15}) once $\sigma(-\infty)=0$ is
imposed as is done in Eq. (\ref{e27}). Actually once normalization
is imposed (improper normalization except at the energy
eigenvalues), there is only one free parameter left. Finally the
quantization condition is obtained by noting that
$\alpha(x\rightarrow \pm\infty)\rightarrow\infty$ (because from
Eqs. (\ref{e27}) and (\ref{e20}) we see that $\alpha^{2}(x)$ is
proportional to $\psi_{1}^{2}+\psi_{3}^{2}$). It then follows from
Eq. (\ref{e24}) that we must have $\sigma(+\infty)=(n+1)\pi$ if
$\psi_{1}$ is to be an eigenfunction and given the symmetry of the
potential, the quantization condition thus reads%
\begin{equation}
\sigma(0)=(n+1)\pi/2, \label{e29}%
\end{equation}
where $n$ is the integer counting the zeros of $\psi_{1}$ on the real line.
Note that when $E$ is not an eigenvalue, Eq. (\ref{e29}) is replaced by%
\begin{equation}
\sigma(0)=(\frac{n}{2}+\frac{1}{4})\pi+\frac{\arctan(2Ic)}{2}. \label{e29b}%
\end{equation}

With regard to the semiclassical limit, we can draw two consequences from Eq.
(\ref{e27}).\ First, by employing directly the asymptotic expansions for
$\psi_{1}$ and $\psi_{3}$, it can be seen that the semiclassical phase
$\sigma^{sc}(x)$ is different from the classical action, except for a single
value of $c$ (see Appendix A). Therefore the classical action $S(x)$, that
appears as the first order solution of Eq. (\ref{e15}) for $R(\xi)=1$ is
indeed a solution of the semiclassical limit of the quantum phase, but one
among others. Second, by employing Eq. (\ref{e13}) in Eq. (\ref{e27}), we
obtain%
\begin{equation}
\tan\sigma(x)=\frac{w_{1}(\xi(x))}{w_{3}(\xi(x);I^{\prime},c^{\prime}))}
\label{e28}%
\end{equation}
where $w_{1}$ and $w_{3}$ are independent solutions of Eq.
(\ref{e13}) lagging $\pi/2$ out of phase. Eq. (\ref{e28})
expresses the phase obtained from Eq. (\ref{e15}) with $R(\xi)=1$
in terms of solutions $\xi(x)$ obeying the phase equation with a
different function $R(\xi)$. Now by replacing $\xi(x)$ by its
asymptotic expansion, we obtain an asymptotic expansion for
$\sigma (x)$.\ For example taking $R(\xi)=\xi$, a solution regular
at $-\infty$ is the Airy function $\mathrm{Ai}(x)$ and an
irregular solution lagging $\pi/2$ out
of phase is $\mathrm{Bi}(x)$, so that to first order we have%
\begin{equation}
\sigma^{sc}(x)=\arctan\frac{\mathrm{Ai}\left[  \xi_{0}(x)\right]
}{\mathrm{Bi}\left[  \xi_{0}(x)\right]  } \label{e30}%
\end{equation}
where $\xi_{0}(x)$ is given by Eq. (\ref{e21}).\ Eq. (\ref{e30})
gives a uniform semiclassical phase $\sigma^{sc}$ (ie, with
$R(\xi)=1$) free of singularities
at the turning points, with $\sigma^{sc}(-\infty)=0,$ $\sigma^{sc}(t_{1}%
)=\pi/6$ and the value $\sigma^{sc}(0)$ defines the quantization condition
when%
\begin{equation}
\arctan\frac{\mathrm{Ai}\left[  \xi_{0}(0)\right]  }{\mathrm{Bi}\left[
\xi_{0}(0)\right]  }=(n+1)\pi/2 \label{e32}%
\end{equation}
where the use of the relevant branch of the $\arctan$ is implicitly
understood. Note that by using the expansions of the Airy functions for large
negative values of the argument,
\begin{equation}
\mathrm{Ai}(-X)=\frac{1}{\sqrt{\pi}\sqrt[4]{X}}\sin\left(  \frac{2}{3}%
X^{3/2}+\frac{\pi}{4}\right)  +O(X^{-7/4})
\end{equation}
for the regular solution and the relevant expansion for $\mathrm{Bi}$, Eq.
(\ref{e30}) takes the familiar approximate form%
\begin{equation}
\sigma^{sc}(x)=S(x)+\pi/4+O(S(x)^{-7/6}). \label{e32z}
\end{equation}
Thus for large values of the action (so in practice for
sufficiently large values of $E$) the semiclassical phase obtained
from appropriate Airy carrier functions is approximately the same,
in the classically allowed region, than the one obtained from the
standard quantization scheme recalled above. However at low
energies the two different expansions of $\sigma^{sc}$ lead to
markedly different phase functions.

\subsection{Conclusion}

Let us recapitulate. First we have recalled that the classical
action is not the only semiclassical phase, nor necessarily the
most useful one in practical computations.\ Certainly, the choice
$R(\xi)=1$ is the most natural one, since it is the only one that
allows to recover the classical action, but we have seen that even
in that case it is possible and advantageous to express the
relevant phase $\sigma(x)$ in terms of alternative phase functions
$\xi(x)$, leading to different semiclassical expansions
$\sigma^{sc}(x)$. Second, quantum phase functions suffer from
ambiguities, because they are irrelevant as quantum mechanical
objects. Even if $R(\xi)=1$ is chosen, there is still a free
parameter (in the form of a boundary condition) that can be varied
at will leading to different behaviour of the phase function. The
phase ambiguity persists in the semiclassical limit: to first
order in $hbar$ there is a single value of the boundary condition
leading to purely classical quantities -- precisely the value that
eliminates remnants of oscillating quantum quantities from the
semiclassical phase (Appendix A).

Now the question can be reversed. Assume that we have chosen the
'right' semiclassical phase, that is the one that only involves
classical quantities. Can the corresponding quantum phase be
constructed? This would amount to
implicitly sum the diverging asymptotic $\hbar$ expansions (\ref{e7})%
-(\ref{e10}). The answer is positive in the case of the harmonic
oscillator, due to the existence of analytic solutions. For other
symmetric oscillators, it does not seem possible to determine this
optimal solution exactly. Rather, Eq. (\ref{e15}) can be solved
(numerically) with a boundary condition allowing to approximate
this optimal solution. Notwithstanding exact quantization can be
achieved and the exact wavefunctions can be retrieved as well.
These points are developped in the subsequent paragraphs of the
paper.

\section{A special case: the harmonic oscillator}

The interest of the harmonic oscillator in the present context
arises from its solvability: analytic solutions can be explicitly
written down and the energy obtained as a function of the
oscillation number in closed form.\ Though the approach given
above is not necessarily useful for solving the harmonic
oscillator problem as such, the harmonic oscillator represents a
system for which the notions introduced in Secs.\ 2 and 3 can be
illustrated on firm grounds. This study will in turn be valuable
for understanding the approximate and numerical treatments that
will be undertaken for general anharmonic oscillators.

\subsection{Analytic solutions and the oscillation number $N(E)$}

Let us solve the Schr\"{o}dinger Eq. (\ref{e4}) with%
\begin{equation}
p^{2}(x)=2\nu+1-x^{2}%
\end{equation}
where we have written
\begin{equation}
E\equiv\nu+1/2\label{e40}%
\end{equation}
so that Eq. (\ref{e4}) takes the form of the Weber equation. A solution
regular at $-\infty$ is given by the parabolic cylinder function which can be
written down in the integral representation as
\begin{equation}
D_{\nu}(z)=\frac{\Gamma(v+1)^{1/2}}{2i\pi}\pi^{-1/4}e^{-z^{2}/2}\oint
e^{-t^{2}/2-\sqrt{2}tz}t^{-v-1}dt
\end{equation}
with the contour encircling the negative real axis and the factors ensure
normalization for the eigenfunctions. A linearly independent solution regular
at $+\infty$ is $D_{\nu}(-z)$ and their Wronskian is seen to be%
\begin{equation}
\mathcal{W}[D_{\nu}(z),D_{\nu}(-z)]=2\pi^{-1}\sin\pi\nu,\label{e42}%
\end{equation}
where we have used $\Gamma(\mu)\Gamma(1-\mu)=\sin\pi\mu/\pi$. Note that the
knowledge of Eq. (\ref{e40}) in conjunction with Eq. (\ref{e42}) suffices to
determine the eigenvalues $E$, since the eigenfunctions are regular at both
$\pm\infty$ and the Wronskian must therefore vanish. By doing so Eq.
(\ref{e40}) is not interpreted as a simple change of variable, but as a
functional relation by which the energy is given in terms of an oscillation
number,
\begin{equation}
N(E)=\nu+1,\label{e43}%
\end{equation}
which gives the number of oscillations of $D_{v}(z)$ between
$x=-\infty$ and $x=\infty$. We shall take for granted that $\nu$
indeed captures the entire oscillatory character of the parabolic
cylinder functions; proofs may be obtained by employing an
asymptotic expansion for $D_{\nu}(z)$ in the interval
$[t_{1},t_{2}]$ \cite{temme00} or by following Olver in
introducing an auxiliary modulus function whose monotonicity on
$[t_{1},0]$ may be proven \cite{olver75}. Note that the first
order semiclassical quantization condition (\ref{e1}) allows to
define a first-order semiclassical oscillation number $N^{sc}(E)$
by the relation
\begin{equation}
N^{sc}(E)=\frac{S(t_{2},E)-S(t_{1},E)}{\pi}+\frac{1}{2},\label{e43b}%
\end{equation}
and the semiclassical quantization condition takes the form
$N^{sc}(E)=n+1$. For the harmonic oscillator the action difference
is $E\pi,$ hence $N^{sc}(E)=E+1/2$, which by Eq. (\ref{e40})
yields the exact quantum relation (\ref{e43}).

\subsection{Quantum and semiclassical phase functions}

\subsubsection{Quantum phases}

We now examine, from the point of view of the formalism introduced in section
3, the choice of the quantum phase.\ We have already seen that working with
circular carrier functions, $R(\xi)=1$, is the most advantageous choice, so
our problem is to find the most relevant phase function $\sigma(x)$ in Eq.
(\ref{e24}). Of course $\psi_{1}$ (by definition regular at $-\infty$; recall
we have imposed $\sigma(-\infty)=0$) is necessarily proportional to $D_{\nu}$,
but there are infinite ways of decomposing $\psi_{1}$ as in Eq. (\ref{e24}).
The main interest in using the amplitude-phase decomposition is probably that
the oscillatory character of the wavefunction is entirely captured in
$\sin\sigma(x)$, so that the amplitude does not oscillate. The unique quantum
phase function displaying such a behaviour will be termed 'optimal'. In this
case, we should have
\begin{equation}
\sigma_{\mathrm{opt}}(\infty)=\pi N(E), \label{e44}%
\end{equation}
(or $\sigma_{\mathrm{opt}}(0)=\pi N(E)/2$ given the symmetry). We
have just seen that the oscillation number is given in the
harmonic oscillator case by $N(E)=\nu+1$. Therefore comparing with
Eq. (\ref{e29b}) gives $c=-\cot\pi \nu/2I$ and comparing Eq.
(\ref{e42}) with Eq. (\ref{e25}) allows to set $I=\pi^{-1}$. These
values of the parameters $I$ and $c$ ensure that the quantum phase
function given by (\ref{e27}) is the optimal one.\ Note that we
must allow for scale transformations
$\psi_{1}\rightarrow\kappa\psi_{1}$ and
$\psi_{2}\rightarrow\kappa\psi_{2}$, implying
$W\rightarrow\kappa^{2}W,$ $I\rightarrow\kappa^{2}I,$
$c\rightarrow c/\kappa^{2}$ so that the parameters corresponding
to the optimal phase actually belong to a class by which $Ic$ and
$W/I$ are given in terms of the oscillation number,
\begin{eqnarray}
Ic  &  =&-\frac{\cot\pi N(E)}{2}\label{e45}\\
\frac{W}{I}  &  =&2\sin\pi N(E). \label{e46}%
\end{eqnarray}

\subsubsection{Semiclassical phases}

The semiclassical phase, defined as the semiclassical limit of $\sigma(x)$ is
given by Eq. (\ref{a4}) in Appendix A. For arbitrary values of $I$ and $c,$ it
is straightforward to see that $\partial_{x}\sigma^{sc}(x)$ will be a highly
oscillating function of $x$. The optimal semiclassical phase is the classical
action, giving a non-oscillating function $\partial_{x}\sigma^{sc}(x)=p(x)$,
obtained for%

\begin{eqnarray}
Ic  &  =-\frac{\cot\left(  S(t_{2})+2\phi\right)  }{2}\label{e48}\\
\frac{W}{I}  &  =2\sin\left(  S(t_{2})+2\phi\right)  . \label{e49}%
\end{eqnarray}

Note that here the set $\{Ic,W/I\}$ refers to the semiclassical
Ermakov system, and is generally different from the quantum set of
parameters; this is why Eq. (\ref{e49}) is verified by
construction. The very special property of the harmonic oscillator
is that the right handside of Eqs. (\ref{e48})-(\ref{e49}) is
equal to the right handside of Eqs. (\ref{e45})-(\ref{e46}), since
to lowest order in $\hbar$ we have $\phi=\pi/4$ and $S(t_{2})$ is
immediately evaluated as $E\pi=\pi(\nu+1/2)$. This property
explains why semiclassical quantization is exact, the mapping
$\{Ic,W/I\}^{\mathrm{sc}}\rightarrow
\{Ic,W/I\}^{\mathrm{quantum}}$ being the identity.

\section{Numerical determination of the quantum phase}

We present in this section an efficient numerical method allowing to compute
the desired quantum phase $\sigma(x).$ This is tantamount to solving directly
Eq. (\ref{e15}) with $R(\xi)=1$.\ From a computational point of view, it is
advantageous to employ a function akin to the Riccati transformed $\zeta$ in
Eq. (\ref{e3}) but incorporating from the start the amplitude-phase
decomposition. Indeed, defining%
\begin{equation}
M(x,E)=\partial_{x}\left[  \sigma(x,E)+\frac{i}{2}\ln(\partial_{x}%
\sigma)\right]  , \label{e50}%
\end{equation}
Eq. (\ref{e15}) becomes equivalent to
\begin{equation}
\partial_{x}M=i\left(  p^{2}(x)-M^{2}(x)\right)  \label{e52}%
\end{equation}
which is a complex but first order nonlinear differential equation. $M$ is
found by an iterative linearization procedure. We introduce the functional%
\begin{equation}
\mathcal{F}(M(x),x)=i\left(  p^{2}(x)-M^{2}(x)\right)  \label{e53}%
\end{equation}
and linearize Eq. (\ref{e52}) by expanding $\mathcal{F}$ to first
order in the vicinity of an initial trial function $M_{0}(x)$. The
resulting first-order linear differential equation is solved for
$M_{1}(x)$, and the process is iterated until convergence is
achieved after $q$ iterations (details are given in Appendix 2).
The exact form of the initial trial function is unimportant (but
see Appendix 2) provided it is smooth and behaves as the converged
solution.\ $M_{0}(x)$ can be conveniently built from the uniform
semiclassical approximation to $\sigma^{sc}(x)$ given by Eq.
(\ref{e30}), since the converged function $M_{q}(x)$ will be close
to the trial function.

The delicate and important point is to determine the boundary condition on Eq.
(\ref{e52}), which holds for all the functions $M_{i}$. It is convenient to
choose the boundary condition at $x=0$: symmetry imposes $\partial_{x}%
\alpha(x=0)=0$ so only the real boundary condition
$\partial_{x}\sigma(x=0)$ needs to be set. Let us take again the
harmonic oscillator as a model. The knowledge of the analytic
solutions and of the oscillation number allow us to determine the
value $\partial_{x}\sigma(x=0)$ for what we called above the
optimal quantum phase. Taking $\psi_{1}(x)=D_{\nu}(x)$, we have
from
Eqs. (\ref{e24}), (\ref{e29b}), (\ref{e43}) and (\ref{e45})%
\begin{equation}
\partial_{x}\sigma(x=0)=\frac{2\left[  \sin\pi N(E)\right]  ^{2}}{\pi\left[
D_{\nu}(0)\right]  ^{2}} \label{e60}%
\end{equation}
where \cite{olver75}%
\begin{equation}
D_{\nu}(0)=\frac{2^{\nu/2}\pi^{1/4}}{\Gamma(\frac{1}{2}-\frac{\nu}{2})\left[
\Gamma(1+\nu)\right]  ^{1/2}}.
\end{equation}
Elementary manipulations on the $\Gamma$ functions yield%
\begin{equation}
\partial_{x}\sigma(x=0)=\frac{2\nu\Gamma(\nu/2)}{(\nu-1)\Gamma(\frac{\nu}%
{2}-\frac{1}{2})}. \label{e62}%
\end{equation}
Employing the boundary condition (\ref{e62}) in the iterative
scheme yields a converged solution $M;$ integrating the real part
of this solution gives us the optimal quantum phase having the
property $\sigma(\infty)=\pi N(E)$ and whose derivatives
$\partial_{x}^{m}\sigma$ are non-oscillating functions. If the
boundary condition is (slightly) different, the quantum phase will
also (slightly) differ from the optimal one: the phase at infinity
will (slightly) differ from the oscillation \ number and (slight)
oscillations will appear in the first derivatives (the
oscillations will only become prominent for high order
derivatives). Recall however that the wavefunctions (\ref{e24})
and (\ref{e26}) are exact solutions of the Schr\"{o}dinger
equation irrespective of the oscillations of the phase-derivative.
Note also that the present numerical scheme does not allow to
construct highly oscillating functions, given that the trial
function is itself not oscillating.

Quantized energies are found by computing
$\sigma(x=\infty,E)=2\sigma(x=0,E)$ as a function of $E$, which in
practice means determining $\sigma(x=\infty,E)$ on a rather loose
energy grid and then interpolate. It is then possible to solve for
the eigenenergies $E_{n}$ by employing the quantization condition
$\sigma(\infty,E)=\pi(n+1),$ which in principle holds irrespective
of the boundary condition. In practice it is however important to
employ adequate boundary conditions so as to keep the
interpolation tractable, which is possible provided
$\sigma(\infty,E)$ is well behaved. Note that the accuracy of the
energy eigenvalues can be improved by tightening the energy grid
in the vicinity of each $E_{n}$; we have generally obtained
eigenvalues with 24 decimal digits without much numerical effort.

\section{Anharmonic oscillators}

\subsection{General setting}

Anharmonic symmetric oscillators are in general not solvable,
except in some exceptional cases (e.g. some states in
shape-invariant potentials in supersymmetric quantum mechanics
\cite{adhikari89}). There are no closed-from solutions to the
Schr\"{o}dinger equation and no such thing as the oscillation
number, or more generally no such thing as a functional relation
$E=f(n^{\ast })$ by which the spectrum is obtained as a function
of a smoothly varying good quantum number $n^{\ast}$.
Semiclassically, such a relation exists by construction, by
inverting $\zeta^{sc}(E)$, $\xi^{sc}(E)$ or $\sigma^{sc}(E)$ (the
absence of the space variable means that the phase has been
integrated on the relevant contour or real line) and quantization
precisely occurs when $n^{\ast}$ is an integer. But we know that
all these semiclassical phase functions are diverging asymptotic
series, and thus the reciprocal relations
$(\zeta^{sc})^{-1}(n^{\ast})$ are only approximate. Employing a
quantum phase represents a compromise.\ It achieves exact
quantization and allows to define relations of the type
$\sigma(E)$ or $\sigma^{-1}(n^{\ast})$ that are exact but
arbitrary (except when $n$ is an integer). As developed in the
preceding sections the treatment is exact and satisfactory from a
quantum-mechanical viewpoint: eigenfunctions and eigenvalues are
determined with a much lower computational cost than the standard
matrix methods, that need to employ large basis. Oscillation
number functions $N(E)$ can be constructed provided that the phase
functions and their derivatives are sufficiently smooth; these
oscillation number functions are approximate and not unique --
they can be made as nearly exact as desired by optimizing the
boundary condition on $M$, but an exact boundary condition of the
type given by (\ref{e62}) in the harmonic oscillator case does not
exist.

The quantum phase employed in the results given below is defined from what is
probably the most intuitive way of setting up the boundary condition: we
determine the value at $x=0$ of the local asymptotic expansion of the phase
derivative (i.e. the inverse of amplitudes),
\begin{equation}
\left.  \partial_{x}\sigma(x)\right|  _{x=0}=\sum_{k=0}^{k\max}\left.
\partial_{x}\sigma_{2k}(x)\right|  _{x=0}\hbar^{2k}; \label{e65}%
\end{equation}
see Eq. (\ref{e16}) with $R(\xi)=1$. Solving the corresponding
recurrence system directly gives us the elements
$\partial_{x}\sigma_{2k}(x)$. $k_{\max}$ is set in principle by
going to the highest possible order before the asymptotic
expansion starts to diverge (so $k_{\max}$ varies with the energy)
and we use Stieljes' simple trick of terminating the series by
multiplying the last retained term by $1/2$. Indeed, the spirit of
the present approach is obtain a quantum phase that would be
'close' in the classically allowed region to the semiclassical
series if the latter converged. Obviously at some point (for large
$k_{\max}$) the extra effort imposed by the computation of high
order terms $\partial_{x}\sigma_{2k}(x)$ is not worth what is
gained by including this term.\ The same can be said about using
super-asymptotic or hyper-asymptotic methods \cite{boyd99}:
getting into involved calculations that would result, say in
changing the 20th decimal number, while not solving in principle
the problem of achieving an optimal boundary condition is probably
not advisable (given, to repeat, that the quantum phase is exact
in all cases).

\begin{figure}[tb]
\includegraphics[height=1.9in,width=2.6in]{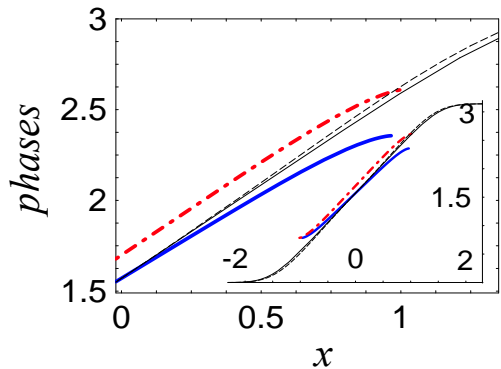}
\caption[]{Quantum and semiclassical phase functions for the
ground state of the homogeneous quartic oscillator. Only the
positive axis is shown (the inset zooms out on the real line). The
quantum phase $\sigma(x)$ (solid black curve) is plotted at the
exactly quantized ground state energy. The classical action
$S(x)+\pi/4$ is shown between the turning points at the WKB
quantized energy (solid thick blue curve) and at the exact
eigen-energy (thick dash-dotted red curve). The $\pi/4$ term is
well-known to be obtained through connection formulae or via an
asymptotic (in $x$) expansion of the uniform semiclassical
approximation for $\sigma$, see Eq. (\ref{e32z}). The first order
uniform semiclassical phase obtained with Airy carrier-functions
is given by the black dashed curve, seen to follow closely the
quantum phase. \label{fq-1}}
\end{figure}

\begin{figure}[tb]
\includegraphics[height=1.9in,width=2.6in]{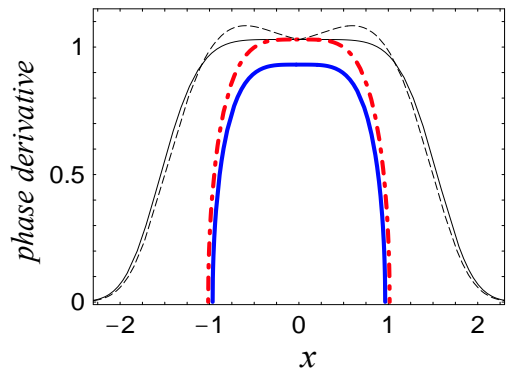}
\caption[]{First derivative $\partial_{x}$ of the quantum and
semiclassical phase functions shown in Fig. \ref{fq-1}.
\label{fq-1-2}}
\end{figure}

\subsection{Results and illustrations}

\subsubsection{The homogeneous quartic oscillator}

The pure quartic oscillator, with the classical momentum function given by%
\begin{equation}
p^{2}(x,E)=2E-x^{4}%
\end{equation}
is the simplest nonsolvable potential, and as such it has been the
object of a great number of works that would be impossible to cite
or summarize. Most of the work however still focuses on the same
topics that motivated the early papers of Bender et al. and Voros,
namely the determination of the numerical properties of the
semiclassical expansion for large orders \cite{bender etal77} and
the development of resummation procedures for the divergent
series, in particular the understanding of analycities that
produce the phenomenon of resurgence \cite{voros83}. Our goal here
is to give a few numerical results so as to show the relevance of
the quantum phase approach.

\begin{figure}[tb]
\includegraphics[height=1.9in,width=2.6in]{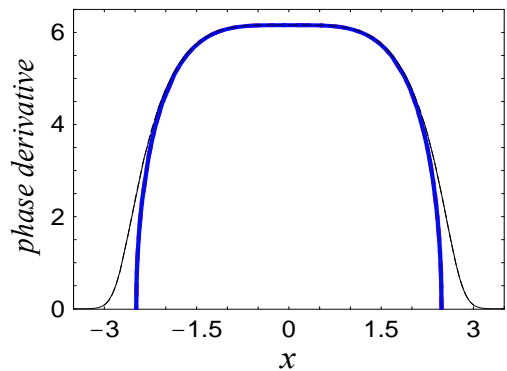}
\caption{{\label{fq-2}}Same as Fig. \ref{fq-1-2} but for the
higher quantized state $n=8$. The two thick curves (solid blue and
dot-dashed red curves) seen in Fig. \ref{fq-1-2} appear now as
superposed given the scale of the plot, as is the case for the two
black curves (the quantum phase function and the uniform
semiclassical one).}
\end{figure}

Fig. \ref{fq-1} shows the quantum phase for the ground state
energy $E_{n=0}=0.53018...$ We also show the WKB phase (the
classical action) calculated at the exact energy and the same
phase determined at the WKB quantized energy
$E_{n=0}^{sc}=0.434...$ (which is off by almost $20\%$) and
finally the first order semiclassical phase obtained with Airy
carrier functions given by Eq. (\ref{e30}) at the corresponding
quantized energy [cf. Eq. (\ref{e32})]
$E_{n=0}^{sc,R=\xi}=0.480...$ Fig. \ref{fq-1-2} shows the
derivatives of these different phases. The following comments can
be made.\ The main observation is that quantum effects are
important: the quantum phase keeps accumulating well beyond the
turning points, in the classically forbidden region. The same
feature is visible for the phase derivative: whereas the classical
momentum vanishes at the turning points, the quantum phase is far
from being negligible even at twice the value of the turning
point. As required, the WKB phase is closer to the quantum one at
the WKB quantized energy, but the phase derivative follows more
closely the quantum curve when taken at the exact quantization
energy. Note that the semiclassical phase obtained with Airy
carrier functions is well behaved through the entire real line and
thus follows the quantum phase in the classically forbidden
regions; the quantum effects in this case show up in the dip
visible in the phase derivative around $x=0$. Note also that this
semiclassical quantization is twice as accurate for the ground
than the first order standard (WKB) semiclassical quantization.
The standard semiclassical quantization scheme can be taken to
higher order: divergence occurs after the third term, so we can go
up to $k_{\max}=3$. In this case, using our simple rule stated
above for the terminant, we find the energy for the ground state
to order $\hbar^{8}$ to be $E_{n=0}^{sc}=0.483...,$ though
stopping the expansion after the second term gives the slightly
better result $E_{n=0}^{sc}=0.490...$. Fig. \ref{fq-2} shows the
situation at a higher energy, for the 8th excited state with the
exact energy $E_{n=8}$ found by solving the quantum phase
quantization condition. We only show the plot for the phase
derivatives because the different phases would barely be
distinguishable on the scale of the plot. The first order standard
semiclassical phase derivatives taken at the exact and at the WKB
quantized energies can not be distinguished on the figure, whereas
the quantum phase and the first order $R(\xi)=\xi$ uniform
semiclassical phase are barely distinguishable on the scale of the
figure. In the classically allowed region the different curves are
very close one to the other and appear as superposed. This trend
is of course expected, given that for symmetric oscillators as $E$
increases the first-order semiclassical approximation improves (in
relative terms).

\begin{figure}[tb]
\includegraphics[height=2.1in,width=4.9in]{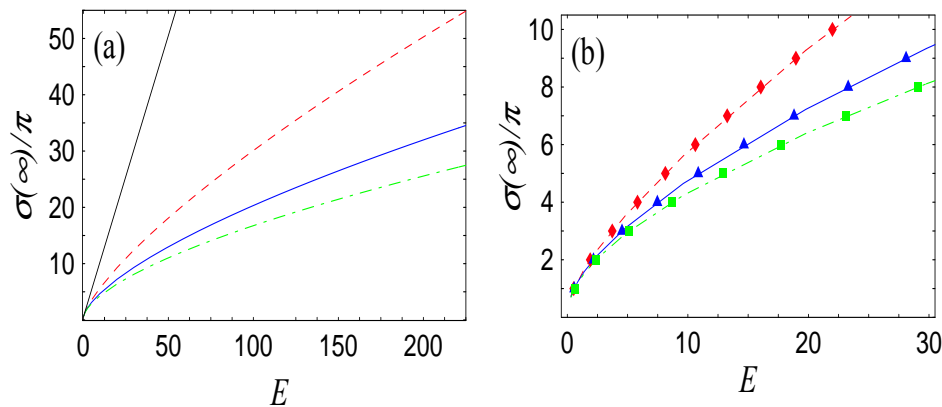}
\caption[]{The total phase $\sigma(x=\infty,E)/\pi$ defining the
oscillation number $\tilde{N}(E)$ is shown for the homogeneous
quartic (dashed red), sextic (solid blue) and octic (dot-dashed
green) oscillators. (a) shows the oscillation numbers up to
$E=220$ a.u.; the straight black line on the left shows $N(E)$ for
the harmonic oscillator. (b) zooms on the lower energy range. The
diamonds, triangles and boxes are plotted each time $\tilde{N}(E)$
is an integer, yielding the exact quantization energies for the
quartic, sextic and octic oscillators respectively. \label{fq-5}}
\end{figure}

Fig. \ref{fq-5} (a) gives $\sigma(x=\infty,E)=2\sigma(x=0,E)$
interpolated as a function of the energy. Fig. \ref{fq-5} (b)
zooms in the lower energy region, the boxes represent the
quantized energies.\ The interest of such a curve is two fold.
First, as already mentioned, we use this interpolated function to
quantize the system with a high numerical precision and a modest
computational cost. Second, this curve defines an approximation to
the oscillation number $N(E),$ introduced above in the context of
the harmonic oscillator. Indeed, $N(E)$ is readily obtained in the
case of solvable potentials, but it is not defined for nonsolvable
problems. The present scheme thus allows to define an approximate
oscillation number to be denoted $\tilde{N}(E),$ counting the
number of half-wavelengths of the wavefunction.

\subsubsection{Sextic and octic oscillators}

\begin{figure}[tb]
\includegraphics[height=1.9in,width=2.6in]{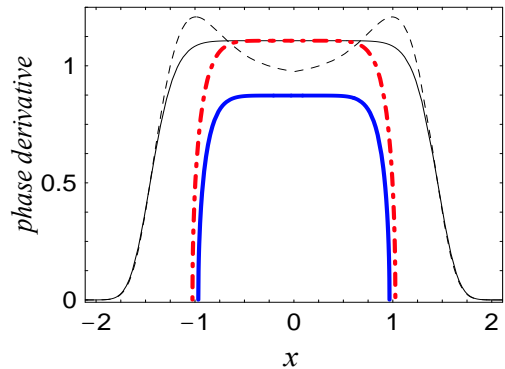}
\caption{{\label{fq-3}}Same as Fig. \ref{fq-1-2} for the ground
state of the pure octic oscillator}
\end{figure}

\begin{figure}[tb]
\includegraphics[height=5in,width=4.in]{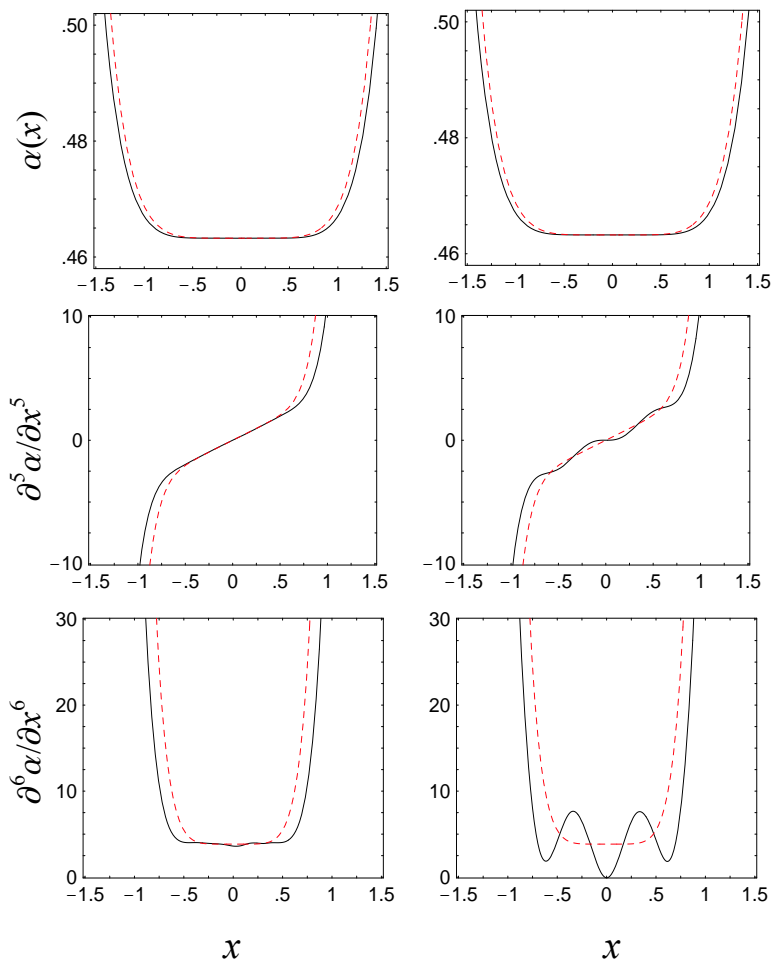}
\caption[]{Amplitude functions and their fifth and sixth
derivatives for the homogeneous sextic oscillator at $E=10.8571$
(near the 4th excited level). The right panel shows $\alpha$,
$\partial _{x}^{5} \alpha$ and $\partial _{x}^{6} \alpha$ for an
amplitude function required to obey the boundary condition
$\alpha^{-2}(x=0)=p(x=0)$. The oscillations indicate that the
amplitude function in this case captures part of the oscillatory
character of the wavefunction \ref{e24}. The left panel shows the
same quantities when the amplitude function obeys the boundary
condition (\ref{e65}) to order 14 in $\hbar$. This amplitude does
not display the oscillations visible in the right panel, although
the boundary conditions differ by less than 3 parts in $10^{-5}$.
To guide the eye, the dashed (red) line represents the
semiclassical amplitude $p(x)^{-1/2}$ (top) and its fifth (middle)
and sixth (bottom) derivatives, identical in the left and right
panels. \label{fq-4}}
\end{figure}

Pure sextic and octic oscillators, ie $V(x)=x^{2m}/2$ with $m=3$
and $4$, have received little attention compared to the
homogeneous quartic case. Indeed, except in the complex trajectory
approach, where the number of turning points in the complex plane
rapidly increases with $m$ \cite{chebotarev99}, all the
homogeneous potentials of higher degree have the same basic
properties that can be seen on the quartic oscillator. In
particular the functions appearing in\ the standard semiclassical
expansion (\ref{e30}) can be obtained in closed form. As $m$
increases, the standard semiclassical quantization procedure
becomes worse (except in the $m\rightarrow\infty$ limit, where a
solvable potential -- the infinite square well is obtained but
then (\ref{e10}) does not apply).\ For example for the ground
state of the octic oscillator, standard semiclassical quantization
breaks down after the crude WKB term and is $38\%$ too low. The
semiclassical quantization condition obtained with Airy carrier
functions is $30\%$ too low, and the dip around the origin is more
severe than in the quartic case, as portrayed in Fig. \ref{fq-3}.\
The oscillation numbers $\tilde{N}(E)$ for the pure sextic and
octic oscillators are displayed in Fig.\ \ref{fq-5}. These
approximate oscillation number functions were obtained by
employing the boundary condition given above. As noted earlier, if
a different boundary condition is set, the resulting phase
function (and thus $\tilde{N}(E)$) will be different. This is
illustrated in Fig. \ref{fq-4} for the amplitude function
$\alpha=(\partial_{x}\sigma)^{-1/2}$ of the sextic oscillator near
the 4th excited level at $E=10.8571$: we have plotted the
amplitude functions
$\alpha(x)$ and the derivatives $\partial_{x}^{5}\alpha$ and $\partial_{x}%
^{6}\alpha$ corresponding to two close but different boundary
conditions. The left panel is obtained with the boundary condition
(\ref{e65}) given above up to $o(h^{16})$ whereas the right panel
corresponds to the 'WKB' boundary condition $\left.
\partial_{x}\sigma(x)\right|  _{x=0}=p(0)$. These two boundary conditions are
very close, less than 3 parts in $10^{-5}$, and $\alpha(x)$ is
indeed seen to be almost identical in both cases. However when
higher derivatives of the amplitude are plotted, differences
appear: the oscillatory structure is already visible in the fifth
derivative of the $p(0)$ amplitude function, leading to radically
different sixth derivatives. This means that when the classical
boundary condition is employed, the quantum amplitude retains part
of the oscillatory character of the wavefunction.

\begin{figure}[tb]
\includegraphics[height=1.9in,width=2.6in]{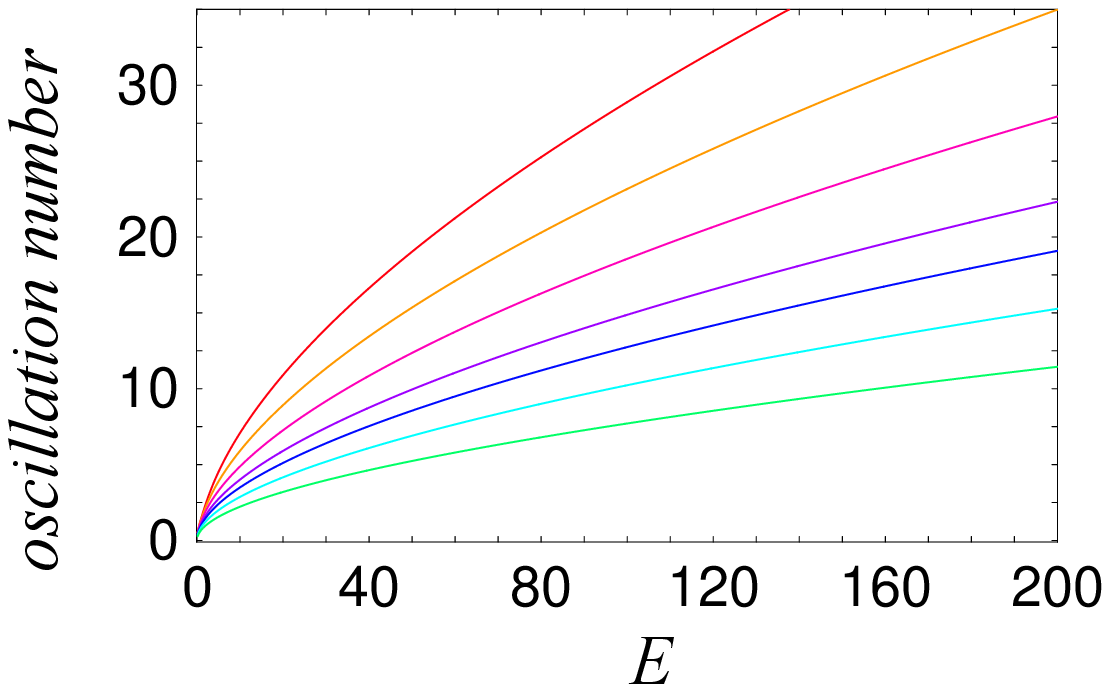}
\caption[]{Oscillation number $\tilde{N}(E,\lambda)$ for the
anharmonic oscillator $x^{2}/2+\lambda x^{10}/2$ with varying
$\lambda$. From top to bottom: $\lambda=0.001$ (red), 0.01
(yellow), 0.1 (pink), 1 (purple), 5 (blue), 50 (light blue) and
1000 (green). \label{fq-7-1}}
\end{figure}

\subsubsection{Anharmonic perturbations}

\begin{figure}[tb]
\includegraphics[height=1.9in,width=2.6in]{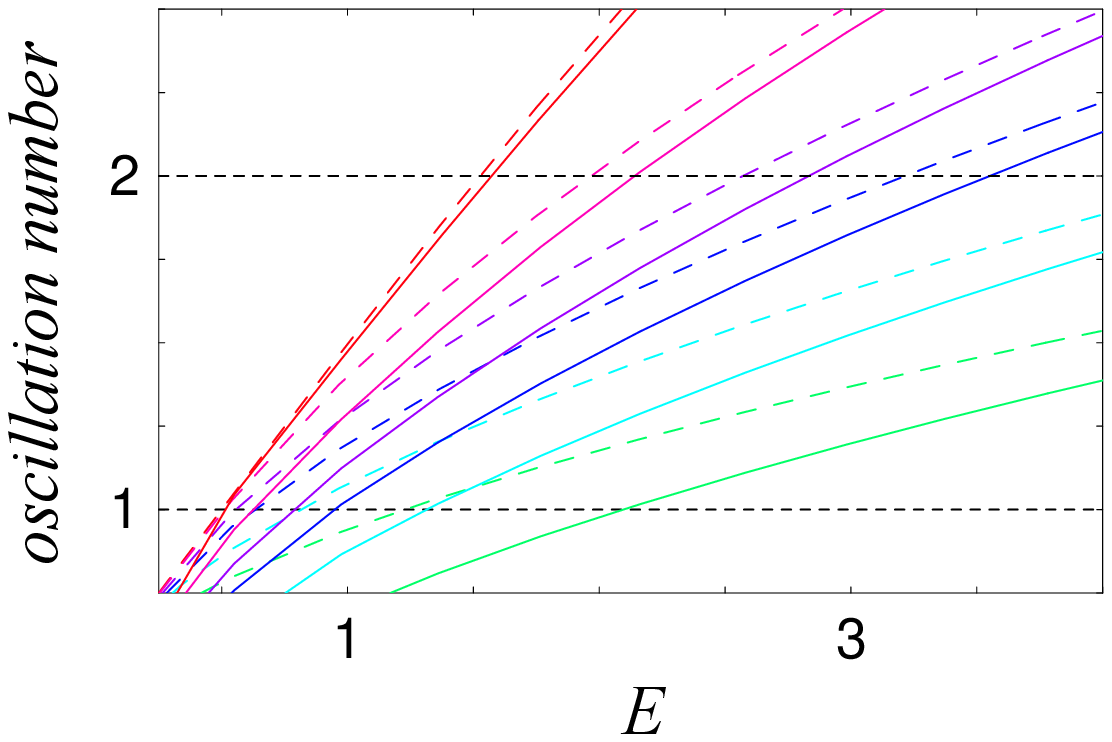}
\caption[]{Quantum $\tilde{N}(E,\lambda)$ (solid lines) and
first-order semiclassical $N^{sc}(E,\lambda)$ (dashed lines)
oscillation number functions for decadic anharmonic oscillators
zoomed at low energy for different values of $\lambda$: from top
to bottom: $\lambda=0.001$ (red), 0.1 (pink), 1 (purple), 5
(blue), 50 (light blue) and 1000 (green). The intersection of the
different curves with the horizontal dashed lines plotted at
integer values of the oscillation number gives the quantization
condition: for example  for $\lambda=1000$ the exact energy of the
ground state ($\tilde{N}(E,\lambda)=1$) is found at $E=2.09$
whereas following the corresponding semiclassical
$N^{sc}(E,\lambda)$ (dashed green line) WKB quantization yields a
ground state energy $E=1.22$ (which turns out to be close to the
exact energy of the $\lambda=50$ oscillator). \label{fq-7-2}}
\end{figure}

\begin{figure}[tb]
\includegraphics[height=2.2in,width=2.5in]{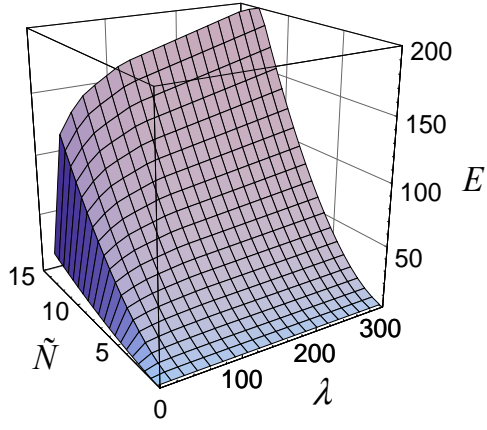}
\caption[]{Correspondence between the oscillation number
$\tilde{N}$ and the energy for decadic anharmonic oscillators as a
function of the coupling constant $\lambda$. \label{fq-8}}
\end{figure}

By far, the main interest in anharmonic oscillators has focused on solving
problems in which anharmonic corrections must be added to the harmonic case.
The simple Hamiltonian%
\begin{equation}
H=\frac{p^{2}}{2}+\frac{x^{2}}{2}+\lambda\frac{x^{4}}{2}%
\end{equation}
in which a quartic perturbation is added to the harmonic
oscillator has given rise to an incredible number of works since
the pioneering paper by Bender and Wu \cite{bender wu69}.\ The two
issues here are the aymptotic semiclassical series, as in the
homogeneous case, but also the behaviour of the eigenvalues
$E(\lambda)$ as a function of the coupling parameter $\lambda.$
The resulting perturbative series is divergent, hence the
development of different resummation schemes based on scaling
transformations or renormalization (eg \cite{skala etal99}). The
summation of the perturbative series for a decadic
perturbation, i.e.%
\begin{equation}
p^{2}(x,E,\lambda)=2E-x^{2}-\lambda x^{10}\label{e70}%
\end{equation}
is considered to be particularly challenging \cite{nunez03}.
Although the method presented in this paper could be
advantageously put to work to study the quantum/classical
correspondence of the perturbative series, and in particular the
behaviour of the different semiclassical expansions introduced
above as a function of $\lambda$, this topic is out of the scope
of this work. We only wish to stress that the method based on the
quantum phase is a powerful tool to investigate the behaviour of
the energies and wavefunctions: rather than having recourse to
different perturbative expansions depending on whether $\lambda$
lies in the weak or strong coupling regimes, the quantum phase
allows to compute essentially exact quantum results within a
unified scheme. It behaves as a semiclassical phase function (in
the sense that physical properties are extracted from the
semiclassical and quantum phase functions in the same way) in
situations in which the semiclassical approximation fails. This is
illustrated in Fig. \ref{fq-7-1}, where we display the oscillation
number $\tilde{N}(E,\lambda)$ for the decadic perturbation
(\ref{e70}) for different values of $\lambda$ ranging from
$\lambda=0.01$ to $\lambda=1000$. Fig. \ref{fq-7-2} zooms on the
lowest states; the broken lines correspond to the semiclassical
oscillation number $N^{sc}(E,\lambda)$ obtained from the classical
action [cf Eq. (\ref{e43b})] for the same values of $\lambda$. The
difference between the quantum and semiclassical curves gives a
measure of the accuracy of the first-order semiclassical
quantization, which as expected gets worse as $\lambda$ increases
and as $E$ decreases. In the strong coupling regime the
semiclassical phase function can barely said to constitute an
approximation: for example the predicted semiclassical energy for
the ground state of the $\lambda=1000$ oscillator is seen to lie
between the exact ground state energies of the $\lambda=5$ and
$\lambda=50$ oscillators; the slopes of the semiclassical and
quantum phase functions, from which periodic time scales are
determined, are markedly different. Finally Fig. \ref{fq-8} gives
the behaviour of the oscillator's energy as a function of
$\lambda$ and $\tilde{N}$. By fixing $\tilde{N}$ at an integer
value $n$ the figure gives the quantization energy of the $n$th
level as a function of $\lambda$. Alternatively by fixing
$\lambda$ we can follow the energy of the system as the
oscillation number increases.

\section{Conclusion}
In summary we have investigated symmetric oscillators, and in
particular their quantization, with the help of semiclassical and
quantum phase functions. Although different semiclassical phase
functions can be defined, the most natural choice corresponds to
the scheme that yields to first order the classical action $S(x)$.
When the classical action and its associated amplitude
$(\partial_{x}S)^{-1/2}$ are plugged into the semiclassical
wavefunction, the entire oscillatory character of the wavefunction
is captured by the oscillations of the phase. We have introduced
quantum phase functions that mimic this property: an arbitrary
quantum phase function would not be useful, but when supplemented
by semiclassical boundary conditions it behaves in a similar way
as the semiclassical phase, allowing to define oscillation number
functions and to retrieve the exact quantum mechanical
wavefunctions and eigenvalues. However the construction in
principle of an 'optimal' phase function, the one that would
generate entirely the oscillations of the wavefunction remained
out of reach: this was possible for the harmonic oscillator case,
thanks to the existence of closed form solutions, but not for the
more general nonsolvable oscillators. The retained solution was to
match the quantum phase to high order pre-divergent semiclassical
expansions in the classically allowed region, so as to obtain
nearly optimal phase functions.

Indeed, our underlying working hypothesis has been that the
optimal quantum phase represents implicitly the resumed divergent
series constituting the semiclassical expansion. It would
therefore be interesting to connect the present approach with the
so-called exact WKB analysis \cite{voros83,voros99}. The
quantizing equation for oscillators in exact WKB analysis arises
from a quantum relation, namely the Wronskian of the function
called $\psi_{2}(x)$ in this paper (appropriately normalized
however) and of $\psi_{2}(q)$ where $q$ represents the rotation of
the real line in the complex plane by a spectral symmetry angle
\cite{voros99}. The Wronskian is determined at $x=\infty$ from an
exact WKB representation of $\psi_{2}$ and at $x=0$ in terms of
spectral determinants. In principle a system of equations that can
be solved numerically can be extracted by equating the expressions
for the Wronskian, but expect for a few particular cases, the
exact WKB analysis as well as methods based on resurgent functions
\cite{pham??} seem to be better qualified to obtain general proofs
rather than numerical results. The method developed in this paper
is some extent the opposite: the QLM method employed to determine
the quantum phase is numerically straightforward and transparent,
but the relation to semiclassics is indirect. The WKB type form of
the initial trial function ensures that the converged result will
not display strong oscillations and the main semiclassical input
enters through the boundary condition. The oscillation number we
define is a rewriting of the Wronskian $W$ of the real solutions
$\psi_{2}$ and $\psi_{1}$ (resp. recessive and dominant at
$\infty$) appropriately renormalised by a factor that takes here
the form of the invariant $I$. The present impossibility of
ascribing a unique value to $W/I$ explains why the oscillation
number, i.e. the total phase accumulated on the real line, retains
some arbitrariness (except at the eigenvalues). Lifting this
arbitrariness should precisely be equivalent to giving a value to
the divergent semiclassical series for the phase.

\appendix
\section{}

In the classically allowed region, let
\begin{equation}
\psi_{1}^{sc}(x)=\left(  \frac{2I}{p(x)}\right)  ^{1/2}\sin\left(
S(x)+\phi\right)  \label{a1}%
\end{equation}
and using the symmetry of the potential%
\begin{equation}
\psi_{2}^{sc}(x)=\left(  \frac{2I}{p(x)}\right)  ^{1/2}\sin\left(
S(t_{2})-S(x)+\phi\right)  ,
\end{equation}
with $\mathcal{W}[\psi_{1},\psi_{2}]=2I\sin\left(  S(t_{2})+2\phi\right)  .$
These representations of the wavefunctions are valid far from the turning
points; we have taken $S(t_{1})=0$ and the connection rules at the turning
point $t_{1}$ impose $\phi=\pi/4$ (although this is not important in this
context). These quantities are substituted into Eq. (\ref{e26}) to obtain
$\psi_{3}^{sc}(x;I,c)$ from which the semiclassical phase $\sigma^{sc}(x)$ is
found as%
\begin{equation}
\cot\sigma^{sc}(x;I,c)=\cot\left(  S(x)+\phi\right)  -\left[  \cot\left(
S(t_{2})+2\phi\right)  +2Ic\right]  . \label{a4}%
\end{equation}
The standard semiclassical result $\sigma^{sc}(x)=S(x)+\phi$ is therefore
obtained when the bracket in the Eq. above vanishes, namely for a single value
of $c$, given as a function of $I$ and $\mathcal{W}$. The same reasoning can
be made for higher order $\hbar$ expansions \cite{matzkin01}. Note that Eq.
(\ref{a4}) for the semiclassical phase involves in the bracket quantum
quantities through the Wronkian and the invariant (these quantities control
the quantization condition and the normalization of the wavefunction).

\section{}

We detail the method employed in Sec.\ 5 to evaluate the quantum phase. The
nonlinear Eq. (\ref{e52}),%
\begin{equation}
\partial_{x}M=\mathcal{F}(M(x),x) \label{a10}%
\end{equation}
where $\mathcal{F}$ is given by Eq. (\ref{e53}) is linearized by expanding the
functional to first order in the vicinity of a function $M_{q}(x)$ which is
almost identical to $M(x)$ within a given preset accuracy (i.e. it represents
the converged answer). This results in the equation%
\begin{equation}
\partial_{x}M=\mathcal{F}(M(x),x)\simeq\mathcal{F}(M_{q}(x),x)+\left.
\frac{\delta\mathcal{F}}{\delta M}\right|  _{M_{q}}\left(  M(x)-M_{q}%
(x)\right)  .
\end{equation}
$\mathcal{F}(M_{q}(x),x)$ is in turn obtained from $M_{q-1}(x)$ by solving the
linear differential equation%
\begin{equation}
\partial_{x}M_{q}=\mathcal{F}(M_{q-1}(x),x)+\left.  \frac{\delta\mathcal{F}%
}{\delta M}\right|  _{M_{q-1}}\left(  M_{q}(x)-M_{q-1}(x)\right)  \label{a12}%
\end{equation}
and so on. Of course in practice the process works the other way round: an
initial trial function $M_{0}(x)$ is chosen and fed into Eq. (\ref{a12}) which
is solved for $M_{q=1}$, which is a better approximation to $M$ than $M_{0}$.
The process is repeated with $q=2$ in Eq. (\ref{a12}), which yields a better
approximation $M_{2}$; the iteration stops when $M_{q}(x)-M_{q-1}(x)$ falls
below the preset precision. This process where the solution of the nonlinear
differential equation (\ref{a10}) is replaced by iteratively solving Eq.
(\ref{a12}) is known as the quasilinearization method (QLM).\ Introduced three
decades ago in the context of linear programming it was first employed in a
quantum mechanical context as an analytical tool to quantize the Coulomb
problem \cite{raghu87}. However the extension of QLM to handle functions with
singularities on unbounded domains, as is the case in quantum mechanics, is
much more recent: Mandelzweig and co-workers \cite{mandelzweig99} proved that
the important property of quadratic convergence which makes this method
powerful was still verified, and developed QLM based numerical tools to solve
the Schr\"{o}dinger equation in several cases of potential scattering and
bound state problems \cite{mandelzweig etal}.

The determination of the trial function seems at first sight
delicate.\ In Ref. \cite{raghu87} $M_{0}(x)=ip(x)$ was taken to be
the most natural choice because it is based on the standard
semiclassical scheme (\ref{e3}), and it was shown that in that
case $M_{1}(x)$ can be obtained as a series of the form $\sum
c_{k}\zeta_{k}^{\prime}(x)$ where the coefficients $c_{k}$ remind
us that $M_{1}$ is only an approximation to the converged solution
$M(x)$. The important point is that taking $M_{0}(x)=ip(x)$ gives
iterated solutions that are singular at the turning points.
Although this property was employed in \cite{raghu87} to quantize
a solvable case (the Coulomb problem) by using an appropriate
contour in the complex plane, it is clearly of limited use in the
more general case of nonsolvable potentials. This is why
implementing the QLM within a real quantum phase approach is
advantageous. $M(x)$ has the form given by Eq. (\ref{e50}), and
the initial trial function $M_{0}$ is sought for
in the form%
\begin{equation}
M_{0}(x,E)=\partial_{x}\left[  \sigma_{0}(x,E)+\frac{i}{2}\ln(\partial
_{x}\sigma_{0})\right]  .
\end{equation}
A good choice for $\sigma_{0}$ is to take the uniform semiclassical phase
$\sigma^{sc}$ given by Eq. (\ref{e30}). This gives a smooth function $M_{0}$
without singularities nor dicontinuities that is known to have the same
behaviour as the converged solution.\ We first thought that $\sigma_{0}%
=\sigma^{sc}$ would the only available choice to implement the QLM with a
quantum phase approach, but the algorithm turned out to be much more flexible.
Indeed, taking%
\begin{equation}
M_{0}(x,E)=\left|  p(x)\right|  \left[  \theta(x-t_{1})\theta(t_{2}%
-x)+i\theta(x-t_{1})-i\theta(x-t_{2})\right]  \label{a15}%
\end{equation}
(or the restriction of this equation to the half-line which is sufficient for
symmetric oscillators) works as well and is faster and easier to implement.
Eq. (\ref{a15}) gives the main components of $M$ in the classically allowed
and forbidden regions: indeed in the forbidden region, $\partial_{x}\sigma$ is
small and tends to $0$ whereas $\partial_{x}\alpha/\alpha\sim\partial_{x}%
\psi_{j}/\psi_{j}\sim\pm i\left|  p\right|  $ where $j=2$ ($1$) in the left
(right) forbidden region, whereas $\partial_{x}\alpha/\alpha$ is small and
$\partial_{x}\sigma\sim p$ in the classically allowed region. The
discontinuities of Eq. (\ref{a15}) at the turning points are smoothed out by
the finite step size employed in the numerical integration routine.


\vspace{1cm}
\begin{thebibliography}{99}                                                                                                %

\bibitem{note1}Solvable refers here to the existence of exact closed form
analytical solutions of the Schroedinger equation.

\bibitem{bender etal77}Bender CM, Olaussen K and Wang PS 1977 Phys. Rev. D
16, 1740.

\bibitem{voros83}Voros A 1983, Ann. Inst. H. Poincare A \textbf{39} 11.

\bibitem{pham??}Delabaere E, Dillinger H and Pham F 1997 J. Math. Phys. 38 6126

\bibitem{balian voros77}Balian R, Parisi G and Voros A 1978 Phys. Rev. Lett.
41 1141

\bibitem{chebotarev99}Chebotarev LV 1999 Ann. Phys. 273, 114

\bibitem{bender wu69}Bender C M and Wu T T 1969 Phys. Rev. 184 1231

\bibitem{slavyanov}Slavyanov S Yu 1996 \emph{Asymptotic Solutions of the
One-Dimensional Schrodinger Equation} (Providence (RI, USA): American
Mathematical Society)

\bibitem{dunham32}Dunham J L 1932 Phys. Rev. 41 713

\bibitem{froman77}Fr\"{o}man N and Fr\"{o}man P O 1977 J Math Phys 18 96

\bibitem{tutik00}Dobrovolsky G A and Tutik R S 2000 J. Phys. A 33 6593

\bibitem{nayfeh}Nayfeh A H 2000 \textit{Perturbation Methods} (New York:
Wiley Classics)



\bibitem{olver}Olver F W J 1974 \textit{Asymptotics and Special Functions}%
(New York: Academic Press)

\bibitem{alvarez00}Alvarez G and Casares C 2000 J. Phys. A 33 2499

\bibitem{froman02}Fr\"{o}man N. and Fr\"{o}man P O 2002  \emph{Physical %
problems solved by the Phase-Integral Method} (Cambridge Univ
Press, Cambridge).


\bibitem{milne-young-wheeler}Milne W E 1930 Phys. Rev. 35 863; Young L A 1931
Phys. Rev. 38 1612; Wheeler J A 1937 Phys. Rev. 52 1123

\bibitem{ray-reid}Reid J\ L\ and Ray J\ R 1980, \textit{J.\ Math. Phys.}
\textbf{21} 1583

\bibitem{matzkin00}Matzkin A 2001 \textit{Phys. Rev. A }\textbf{63} 012103

\bibitem{matzkin01}Matzkin A 2001 J Phys A 34 7833

\bibitem{thylwe05}Thylwe K E 2005 J Phys A 38 235

\bibitem{temme00}Temme N M 2000 J Comp App Math 121 221

\bibitem{olver75}Olver F W J 1975 Phil. Tr. Roy. Soc. A 278 137

\bibitem{adhikari89}Adhikari R, Dutt R and Varshni Y P 1989 Phys Lett A 141 1

\bibitem{boyd99}Boyd J P 1999 Acta Appl. Math. 56, 1

\bibitem{skala etal99}Skala L, Cizek J and Zamastil J 1999 \textit{J. Phys. A
}\textbf{32} 5715

\bibitem{nunez03}Nu\~{n}ez M A 2003 Phys Rev E 68 016703

\bibitem{voros99}Voros A 1999 J Phys A 32 5993

\bibitem{raghu87}Raghunathan K and Vasudevan R 1987 J. Phys. A 20 839

\bibitem{mandelzweig99}Mandelzweig V B 1999, J. Math. Phys. 40 6266

\bibitem{mandelzweig etal}Krivec R and Mandelzweig V B 2003 Comput. Phys.
Comm. 152 165; Krivec R, Mandelzweig V B and Tabakin F 2004 Few-Body Syst 34 57
\end{thebibliography}
\end{document}